\documentclass[preprint,prd,showpacs,preprintnumbers,amsmath,amssymb]{revtex4}

\def \beq {\begin{equation}}
\def \eeq {\end{equation}}
\def \beqar {\begin{eqnarray}}
\def \eeqar {\end{eqnarray}}
\def \bsubeq {\begin{subequations}}
\def \esubeq {\end{subequations}}
\def \beqaw {\begin{eqnarray*}}
\def \eeqaw {\end{eqnarray*}}
\def \pa {\partial}
\def \eps {\varepsilon}
\def \d {\,\mathrm{d}}

\begin{document}


\title{Relativistic Kinetics of Phonon Gas in Superfluids}%
\author{Vladimir Popov}%
\email{vladimir.popov@ksu.ru}%
\affiliation{Department of General Relativity and Gravitation,
Kazan State University, 420008 Kazan, Russia}

\date{\today}%

\begin{abstract}
The relativistic kinetic theory of the phonon gas in superfluids
is developed. The technique of the derivation of macroscopic
balance equations from microscopic equations of motion for
individual particles is applied to an ensemble of quasi-particles.
The necessary expressions are constructed in terms of a Hamilton
function of a (quasi-)particle. A phonon contribution into
superfluid dynamic parameters is obtained from energy-momentum
balance equations for the phonon gas together with the
conservation law for superfluids as a whole. Relations between
dynamic flows being in agreement with results of relativistic
hydrodynamic consideration are found. Based on the kinetic
approach a problem of relativistic variation of the speed of sound
under phonon influence at low temperature is solved.
\end{abstract}


\maketitle

\section{\label{intro}Introduction}

Over the last years the problems of relativistic superfluid
dynamics were discussed repeatedly. Such an interest is well
justified. Superfluids turned out to be the object that has
revealed some problems of relativistic hydrodynamics and
thermodynamics in a new fashion. This has led to earnest
discussion in a relativistic generalization of Landau two-fluid
model \cite{Israel,KhalatLeb,LebKhal,KhalatCarter1,Carter2}.

Usually, two approaches are used in relativistic hydrodynamics,
Eckart and Landau-Lifschits ones. In the former the macroscopic
velocity of a gas (or liquid) is defined as the unit vector
parallel to the particle number flow, and in the latter it does as
eigenvector of the energy-momentum tensor. In equilibrium all
flows such as a particle number flow and an entropy flow are
parallel to this unit vector. When dissipation effects are taken
into account all the macroscopic values get additions that are
orthogonal to the vector of the macroscopic velocity.

Superfluids fundamentally differs from the perfect fluid. The
particle number flow $n^i$ and the entropy flow $S^i$ as well as
conjugated dynamic $\mu_i$ and thermal $\Theta_i$ momenta have
different directions even in equilibrium. It allows to develop an
approach that does not prefer any direction. In context of this
approach it becomes possible not only to describe adequately
relativistic superfluid dynamics but also to discuss some problems
of the traditional approaches \cite{Carter2,Carter6}.

According to \cite{Carter2, Carter6} the energy-momentum tensor
of superfluids at non-zero temperature can be represented in
the form
\beq \label{Tij(SF)}
\mathcal{T}^i_j= n^i \mu_j + S^i \Theta_j - P \delta^i_j,
\eeq
where  $P$ is the liquid pressure. Though the notation
(\ref{Tij(SF)}) does not demonstrate the symmetry of the tensor in
an explicit form, it really takes place \cite{Carter6, Carter3}.

Among the four flows, $n^i, S^i, \mu^i$, and $\Theta^i$, only two,
$n^i$ and $S^i$ say, are independent. The other two are expressed as
\beq                   \label{ABCcoef}
\mu^i=B n^i +A S^i, \quad \Theta^i = A n^i +C S^i,
\eeq
with three scalar function $A, B$, and $C$ named the anomaly, bulk,
and caloric coefficients respectively.
Consideration of superfluidity
supposes the coefficient $A$ should be negative and the coefficients
$B$ and $C$ should be positive \cite{Carter2}.

Since there is no a preferred flow in the theory developed in
\cite{Carter2, Carter6}  the vector of the ``hydrodynamic''
velocity may be associated with any of them. The choice is made by
convenience reason. For example, the velocity may be parallel to
the particle number flow when a gas of identical particles is
considered, whereas it may be direct along entropy flow if
thermodynamic properties are studied. In this work there also
exists a preferred vector $V^i$. It has a sense of a superfluid
velocity. In context of this paper the superfluid velocity plays a
role of a velocity of the carrying medium where quasi-particles,
phonons and rotons,  are located.

Relativistic superfluid hydrodynamics can be used also for solving
more practical problems in different areas apart from studying fundamental
matter. This formalism is found to be efficient to describe
processes in neutron stars
\cite{Comer,Lindblom,Yakovlev1,AnderComer1} %
and to construct the cosmological models
\cite{Volovik2,Volovik3,Visser}.

Hydrodynamic approach is essentially macroscopic and therefore it
does not take into account microscopic processes that have
influence on the dynamic and thermal parameters (such as entropy,
pressure etc.). It is also clear that hydrodynamics can not
describe all the phenomena in superfluids. For example, usage of
hydrodynamic equations is restricted by the condition
$\omega\tau\ll 1$, where $\omega$ is the sound frequency, $\tau$
is the free path time. This condition is not fulfilled at
temperatures close to zero because free path time of the
quasi-particles increases. In this temperature range system
behavior is described by a kinetic equation.

Meanwhile kinetic phenomena in relativistic superfluids remain
beyond the scope of study. The superfluid kinetic theory concerns
the problems related to the excitation gas, which significantly
determines values of fluid parameters at non-zero
temperature. When dissipation is discounted, the quasi-particle gas
determines thermal effects in the first place. In hydrodynamics
influence of the quasi-particles is masked by macroscopic flows
such as an entropy flow. In essence the unit vector in the
direction of the entropy flow is considered as a macroscopic
velocity of the quasi-particle gas.

The kinetic theory of the quasi-particle gas for non-relativistic
superfluids was developed in~\cite{Khalat}. To generalize this
theory for the relativistic case is the primary purpose of this
study. Covariant kinetic theory of the quasi-particles naturally
develops and supplements relativistic kinetic theory for the real
mass and massless particles.

One of the main goals of kinetic theory is to connect macroscopic
variables with intrinsic microscopic parameters. With respect to
superfluids this means that a quasi-particle contribution into the
superfluid dynamic parameters should be obtained. At the low
temperatures we can take into account only phonons (i.e.
excitations with the linear dependence between the energy and the
momentum) as influence of other quasi-particles may be neglected.

When we deal with quasi-particles in special or general
relativity, a main difficulty is to define  the macroscopic
variables adequately. The usual definitions \cite{deGroot} are
inapplicable here, since the quasi-particles being the massless
move slower than light. In Sec.~\ref{macro} this problem is solved
using the Hamilton function for the quasi-particles. Structure of
the function is based on the acoustic metrics introduced in
Sec.~\ref{ametrics}. This permits to modify the usual
constructions of the energy-momentum tensor, the particle number
flow and the entropy flow. In Sec.~\ref{PhononContrib} balance
equations for the quasi-particle gas are derived from the kinetic
equations. Simultaneous consideration of the phonon balance
equation and conservation laws for all liquid allows to find a
phonon contribution to the chemical potential and the pressure of
the liquid. In Sec.~\ref{sound}  the problem of a sound speed
variation under influence of the phonon gas at low temperature is
investigated for the Minkowski and Robertson-Walker metrics.

\section{\label{ametrics}Acoustic metrics}
In the classical Landau's theory \cite{KhalatMono, LL} a
superfluid state is governed by quasi-particles. Each
quasi-particle has a certain momentum and energy that are
functionally related. Such a functional dependence is said to be
an energy spectrum of superfluids. In the spectrum of liquid
helium one recognizes two kinds of quasi-particles, named phonons
and rotons that contribute maximally to thermodynamic properties.
Phonons correspond to the linear part of the spectrum whereas
rotons belong to the nonlinear part of the spectrum. Processes
being due to rotons are inessential at a temperature close to zero
and their contribution is neglected. Thus at the low temperature a
quasi-particle gas may be considered as a purely phonon gas.

To develop our theory it is convenient to introduce so-called
acoustic or phonon metrics \cite{Sing,unruh}
\beq \label{phonon metric}
\overline{g}^{ij}=g^{ij}+\left( \frac{c^2}{c_p^2}-1\right)V^iV^j,
\eeq
where $V^i$ is the superfluid velocity normalized by $V_iV^i=1$,
and the sound speed $c_p$
at zero temperature is defined by the relation
\beq    \label{def Cp}
\frac{c_p^2}{c^2}=\frac{\d P_0}{\d \rho}=\frac{n_{||}}{\mu_0}\frac{\d\mu_0}{\d n_{||}},
\eeq
in which $n_{||}\equiv n^iV_i$, $n^i$ is the \emph{liquid}
particle number flow, $\mu_0$ is the chemical potential, $P_0$ is
the pressure, and $\rho$ is the energy density of the liquid at
zero temperature \cite{FirstSound}. The acoustic metrics is
constructed so that phonon momenta satisfy the relation
\beq           \label{phonon}
\overline{g}^{ij}p_ip_j=0,
\eeq
i.e.~they are tangent vectors to the characteristic hypersurface
of sound propagation in the medium. Along with the tensor
$\overline{g}^{ij}$ the metric tensor with subscripts is involved
\beq
\overline{g}_{ij}=g_{ij}+\left( \frac{c_p^2}{c^2}-1\right)V_iV_j,
\eeq
which satisfy the condition
$\overline{g}^{ij}\overline{g}_{jk}=\delta^i_k$.

Notice that the acoustic metrics (\ref{phonon metric}) can be
formally derived as the special case of a broader class of a
tensor corresponding to an arbitrary energy spectrum. Actually, we
can define the tensor
\[
G^{ij}=g^{ij}+\left(\frac{\pi^2}{\eps^2}-1\right)V^iV^j,
\]
where $\eps=p_iV^i$ is the energy, and $\pi^2=-\Delta^{ij}p_ip_j
\equiv -(g^{ij}-V^i V^j)p_ip_j$ is the square of 3-momentum of a
quasi-particle in the local rest frame of the medium. In this case
$G^{ij}p_ip_j\equiv 0$. If we will establish a functional
dependence between $\eps$ and $\pi$ we obtain the covariant form
of the energy spectrum. However in general, $G^{ij}$ can not be
used as a metrics, since it depends not on only coordinates but
also on momenta. Only for phonons with the condition
\beq   \label{DispPhon}
\eps=\frac{c_p}{c}\pi,
\eeq
$G^{ij}$ can be considered as a metrics.

The superfluid velocity appearing in the acoustic metrics sets the
direction for the ``superfluid momentum'' $\mu_j=\mu V_j$, where
$\mu$ is the chemical potential of superfluids \cite{Carter2}. The
momentum $\mu_j$ is gradient of a scalar function (which is
interpreted as a phase of a wave function in a quantum description
of superfluidity): $\mu_j=\nabla\!_j\phi$. This permits to derive
the equation
\beq \label{sf cond}
\nabla_{[i}\mu_{j]}=0,
\eeq
that demonstrates irrotational nature of superfluid moving.
Contraction of (\ref{sf cond}) with $V^i$ gives the relation
\beq \label{potenV}
\mu D V_j=\Delta_j^i\nabla\!_i \mu,
\eeq
($D\equiv V^i\nabla_i$ is the convective derivative) that can be
interpreted as generalization of the nonrelativistic equation for
the superfluid velocity. The irrotational condition can be
obtained in a more traditional form by convolution of (\ref{sf
cond}) with the projector $\Delta_k^i$ and using (\ref{potenV}):
\beq \label{rotV=0}
\Delta_{i[k} \nabla^i V_{j]}=0.
\eeq

\section{\label{macro}Macroscopic variables}
To describe the phonon gas it is necessary to introduce the usual set
of macroscopic
parameters, such as particle and entropy flow four-vectors
and an energy-momentum tensor, which are constructed using
a phonon distribution
function $f(x,p)$. If to follow the usual way \cite{deGroot, Sing}
the macroscopic variables should be defined as
\bsubeq    \label{old defs}
\beqar
N^i \!\!&=& \!\!2c \! \int \! p^if(x,p)
\frac{\Theta(p^0)\delta(\overline{g}^{ij}p_ip_j)}{(2\pi\hbar)^3}
\frac{\d^4p}{\sqrt{-g}},
                                           \label{old Ni}\\
T^{ij} \!\!&=& \!\!2c \! \int \! p^jp^if(x,p)
\frac{\Theta(p^0)\delta(\overline{g}^{ij}p_ip_j)}{(2\pi\hbar)^3}
\frac{\d^4p}{\sqrt{-g}},
                                            \label{old Tij}
\eeqar
\esubeq
but these expressions are not applicable in this case since they
do not carry required physical content. To illustrate this, let us
consider the vector $N^i$ in the flat space-time.

The component $N^0$ must have meaning of a number particle
density, i.e.
\beq      \label{N0}
N^0=c \int f(x,p)\d^3p,
\eeq
where the integration is to be taken over usual (three-dimensional)
momentum space and $p_0$ is determined by (\ref{phonon}).
$N^\alpha$ ($\alpha=1,2,3$) are the components of a particle flow
three-vector. Therefore in the medium rest frame the phonon particle
flow should be defined as
\beq      \label{Nalpha}
N^\alpha=c_p \int \frac{p^\alpha}{p}f(x,p)\d^3p.
\eeq

$\delta$-function provides transformation of (\ref{old Ni}) to
the usual momentum space
\beq      \label{Ni3D}
N^i=c \int \frac{g^{ij}p_j}{\overline{g}^{0k}p_k}f(x,p)\d^3p.
\eeq
Comparing this expression with (\ref{N0}), one can conclude
that $N^i$ defined according (\ref{old Ni}) can not be used
as a particle number flow.

To give a physical sense to the macroscopic currents, their
definitions should be modified. For this purpose we introduce
Hamilton function for a quasi-particle
\beq                \label{Hamilton}
\mathcal{H}=\frac{1}{2}\overline{g}^{ij}p_ip_j,
\eeq
that is equal to zero at the hypersurface defined by the
expression (\ref{phonon}). Using this function we construct
macroscopic variables
\bsubeq     \label{new defs}
\beqar
N^i &=& c\int \frac{\pa\mathcal{H}}{\pa p_i}f(x,p)
\frac{\Theta(p_jV^j)\delta(\mathcal{H})}{(2\pi\hbar)^3}
\frac{\d^4p}{\sqrt{-g}},
                                            \label{def Ni}\\
T^i_j &=& c\int p_j\frac{\pa\mathcal{H}}{\pa p_i}f(x,p)
\frac{\Theta(p_jV^j)\delta(\mathcal{H})}{(2\pi\hbar)^3}
\frac{\d^4p}{\sqrt{-g}},
                                             \label{def Tij}\\
S^i &=& -c k_B \int \frac{\pa\mathcal{H}}{\pa p_i}[f\ln f-
\nonumber\\&&\!\!\!\!
(1+f)\ln (1+f)]
\frac{\Theta(p_jV^j)\delta(\mathcal{H})}{(2\pi\hbar)^3}
\frac{\d^4p}{\sqrt{-g}}.
                                            \label{def Si}
\eeqar
\esubeq
Further we will use a compact notation
\[
\d \mathcal{P}
=\Theta(\eps)\delta(\mathcal{H})\d^4p/(2\pi\hbar)^3\sqrt{-g}.
\]

The macroscopic flows defined in (\ref{new defs}) satisfy all the
wished requirements. To verify this fact let us consider the
modified vector $N^i$. Integration of (\ref{def Ni}) with
$\delta$-function leads to the expression (\ref{Ni3D}), but now
the acoustic metrics $\overline{g}^{ij}$ (not $g^{ij}$) is in the
numerator, providing fulfilment both (\ref{N0}) and
(\ref{Nalpha}).

The determinations (\ref{old defs}) usually include $\Theta(p^0)$
that gives the positivity of the particle energy. When we deal
with quasi-particles, a role of the energy is assigned to the
convolution $p_iV^i\equiv\eps$. That is why it fill the place of
the argument to the $\Theta$-function in the expressions (\ref{new
defs}).

The expressions (\ref{new defs}) can be considered as
generalization of the traditional definitions of the macroscopic
flows. To revert to the usual notations, Hamilton function should
be taken in the form
\[
\mathcal{H}=\frac{1}{2} [ g^{ij} p_i p_j - m^2 c^2 ]
\]
for massive particles, and
\[
\mathcal{H}=\frac{1}{2} g^{ij} p_i p_j
\]
for massless ones.

The entropy flow definition may be different according to particle
statistics \cite{Stewart}. Phonons in superfluids are bosons, that
leads to the construction in the square brackets of (\ref{def
Si}).

The energy-momentum tensor defined by (\ref{def Tij}) is not
symmetric in the general case. This implies that the angular
momentum of the phonon gas is not conserved \cite{LL2}. This is a
quite expected circumstance because the gas is not an isolated
system. The liquid is not only a medium for phonons, its
parameters are influenced by the quasi-particles. In view of this
interrelation, the energy-momentum of the phonon gas is also not
conserved as we shall see later. Conservation laws hold for the
system as a whole. It will allow us to make conclusion about a
phonon contribution in the thermodynamic parameters of
superfluids. The angular momentum for all the system is implied to
be conserved, therefore the total energy-momentum tensor can be
represented in the symmetric form \cite{Carter6,LL2}.

The macroscopic flows (\ref{new defs}) can be naturally split into
contributions parallel and orthogonal to $V^i$. It is convenient
to decompose the phonon energy-momentum tensor as following:
\beq
\label{decomposition} T^i_j=wV^iV_j+Q^iV_j+Q_jV^i+\Pi^i_j+ \left(
\frac{c^2}{c_p^2}-1 \right)Q_jV^i, \eeq where \beqaw &&
w=T^i_jV_iV^j=\frac{c^3}{c_p^2}\int\eps^2f(x,p)\d \mathcal{P},
\\&&
Q_i=T^k_jV^j\Delta_{ki}=c\int\eps \pi_if(x,p)\d \mathcal{P},
\\&&
\Pi^i_j=T^k_l\Delta_k^i\Delta_j^l=c\int\pi^i \pi_jf(x,p)\d
\mathcal{P},
\\&&
\eps=p_iV^i, \qquad \pi_i=\Delta^k_i p_k.
\eeqaw
A similar decomposition one can perform for the entropy flow:
\beq   \label{decomposeS}
S^i=s_{||}V^i+\sigma^i,
\quad s_{||}=S^iV_i, \quad \sigma^i=\Delta^i_j S^j.
\eeq

\section{\label{PhononContrib}Phonon contribution into superfluid dynamics}
\subsection{\label{KinEq}Kinetic equation for phonons}
The phonon distribution function satisfy Liouville equation
\cite{Klim, Vlasov}
\beq              \label{Loivill}
\frac{\d f}{\d s}=\frac{\pa f}{\pa x^i}\frac{\pa \mathcal{H}}{\pa p_i}-
                \frac{\pa f}{\pa p_i}\frac{\pa \mathcal{H}}{\pa x^i}
               =J(x,f),
\eeq
where canonical Hamilton equations are used
\beq
\frac{\d x^i}{\d s}=\frac{\pa \mathcal{H}}{\pa p_i},\qquad
\frac{\d p_i}{\d s}=-\frac{\pa \mathcal{H}}{\pa x^i},
\eeq
and $J(x,f)$ is a collision integral for the phonons.

The derivative of the Hamilton function (\ref{Hamilton})
with respect to coordinates
can be converted:
\beqar
\frac{\pa \mathcal{H}}{\pa x^i} &=& \frac{1}{2}p_k p_l
\frac{\pa\overline{g}^{kl}}{\pa x^i}=
-p_k p_s \overline{g}^{sm} \overline{\Gamma}^k_{im} =
\nonumber\\&& -\overline{\Gamma}^k_{im} p_k \frac{\pa
\mathcal{H}}{\pa p_m}.
\eeqar
Then the equation (\ref{Loivill}) takes the form:
\beq \label{CartanLiouville}
\frac{\d f}{\d s}
=\frac{\pa \mathcal{H}}{\pa p_i} \widehat{\nabla}_i f=J(x,p),
\eeq
where
\beq
\widehat{\nabla}_i = \frac{\pa }{\pa
x^i}+\overline{\Gamma}^k_{im} p_k \frac{\pa }{\pa p_m}
\eeq
is the Cartan derivative \cite{Vlasov} and
$\overline{\Gamma}^m_{ij}$ are the Christoffel symbols defined
with respect to the metrics $\overline{g}^{ij}$.

One can directly verify that the following equalities are valid:
\beqar
&\displaystyle\widehat{\nabla}_i p_k=0,&\\
&\displaystyle\widehat{\nabla}_i \frac{\pa \mathcal{H}}{\pa
p_k}=0, \eeqar as well as \beq \frac{1}{\sqrt{-\overline{g}}}
\frac{\pa\sqrt{-\overline{g}}}{\pa x^i}=
\frac{1}{\sqrt{-g}}\frac{\pa\sqrt{-g}}{\pa x^i}+ \frac{1}{c_p}
\frac{\pa c_p}{\pa x^i}.
\eeq
Taking into account these properties it is easy to show that
\begin{widetext}
\beq
\int\frac{\pa \mathcal{H}}{\pa p_k} p_{i_1} .. p_{i_l}
\widehat{\nabla}_k f(x,p) \d \mathcal{P} = \overline{\nabla}_k
\int \frac{\pa \mathcal{H}}{\pa p_k} p_{i_1} .. p_{i_l} f(x,p)\d
\mathcal{P}- \frac{1}{c_p} \frac{\pa c_p}{\pa x^k} \int \frac{\pa
\mathcal{H}}{\pa p_k} p_{i_1} .. p_{i_l} f(x,p)\d \mathcal{P},
\label{Relat1}
\eeq
\end{widetext}
where $\overline{\nabla}_k$ is covariant derivative
defined with respect to the metrics
$\overline{g}_{ij}$.

\subsection{\label{BalanceEqs}Balance equations of phonon gas}
The relation (\ref{Relat1}) is very convenient to derive
macroscopic equations. Let us multiply the Liouville equation
(\ref{CartanLiouville}) on $p_{i_1} .. p_{i_l}$ and integrate it
over momenta. If number of the momenta $l=0$ then we obtain the
balance equation \beq  \label{balanceNph} \nabla_i N^i = \int
J(x,p)\d P. \eeq Notice that the covariant derivative is
associated with the metrics $g^{ij}$ and a rate of change of the
quasi-particles is governed by the collision term of the kinetic
equation. Therefore the equation (\ref{balanceNph}) becomes a
conservation law only in equilibrium when $J(x,p)=0$.

There is a different situation for phonon energy-momentum balance,
that can be obtained setting $l=1$. In this case the integral of
the right hand side of the Liouville equation turns to zero by a
momentum conservation law and hence we find that
\beq \label{balance Tij}
\nabla_i T^i_j+(\Gamma^m_{ij}-\overline{\Gamma}^m_{ij})T^i_m=0,
\eeq
where $\Gamma^m_{ij}$ and $\overline{\Gamma}^m_{ij}$ are the
Christoffel symbols with respect to the metrics $g_{ij}$ and
$\overline{g}_{ij}$ correspondingly. The second term in
(\ref{balance Tij}) is found using the explicit expression for
$\overline{g}_{ij}$ and decomposition (\ref{decomposition}). Then
the balance equation (\ref{balance Tij}) can be written as
\beq \label{SaveTij}
\nabla_i T^i_j+\left( \frac{c^2}{c_p^2}-1 \right)Q^i\nabla_jV_i -
\frac{w}{c_p} \nabla_j c_p =0.
\eeq
This means that there is no a phonon energy-momentum conservation even
in equilibrium.

Contracting the equation (\ref{SaveTij}) with $V^j$ and applying
the decomposition (\ref{decomposition}), we find
\beq
\label{convolVdT}
Dw+w\theta +\nabla_i Q^i-Q^i DV_i
-\Pi^{ij}\nabla_iV_j- \frac{w}{ c_p} D c_p =0,
\eeq
where the
following notations are introduced:
\beq   \label{peripherals}
\theta=\nabla_iV^i=g^{ij}\theta_{ij}=\Delta^{ij}\theta_{ij},
\quad \theta_{ij}=\Delta_{k(i}\nabla^k V_{j)}.
\eeq

Next we follow Carter \cite{Carter6}, who noticed that there is no
evidence of parallelism of entropy and particle flows in the
hydrodynamic equations even in equilibrium. Formally we may
construct hydrodynamics with unparallel equilibrium flows. Really
it is realized in superfluid hydrodynamics. In usual liquid the
particle and entropy flows become unparallel in non-equilibrium
only, but traditionally a basic thermodynamic relations (that are
related to a rest frame of the velocity selected) remain the same.
Such an approach seems to be not well consistent since the flows
have different directions, and hence, one more independent
parameter appears in addition to a usual set of variables. It is
an absolute value of difference of the flows. Hence the Gibbs
relation have to be supplemented by a term of a kind $X\delta Y$,
where $Y$ is a new variable and $X$ is a thermodynamically
conjugated current. Using a Legendre type transformation the new
parameters may be added in the Gibbs relation as $-Y\delta X$. If
$Y=\sqrt{y_iy^i}$ then the variation of the scalar can be
transformed to the variation of the vector: $X\delta
Y=(Xy_i/Y)\delta y^i=x_i\delta y^i$.

In our consideration the superfluid velocity and entropy flow have
different directions and the Gibbs relation will be used in the
following way
\beq \label{Hibbs}
\delta w =T\delta s_{||}+\sigma^i \delta \tau_i +
\mu_1\delta n_{||},
\eeq
where $T$ is the temperature, $\tau_i$ is the current conjugated
to $\sigma^i$. The third term appears because the phonon energy
depends on the liquid particle density. The parameter $\mu_1$ is
not the chemical potential of the phonon gas (it equals to zero
since the number of quasi-particles is not conserved) and it
should be determined as well as $\tau_i$.

Taking into account (\ref{decomposeS}) the entropy production density
can be realized in the form
\beq \label{SProd}
\nabla_i S^i =Ds_{||}+s_{||}\theta+\nabla_i \sigma^i.
\eeq
The liquid particle number flow
can be decomposed at components parallel and
orthogonal to  the superfluid velocity
\beq   \label{decomposeN}
n^i=n_{||}V^i+H^i,
\qquad n_{||}=n^iV_i, \qquad H^i=\Delta^i_j n^j.
\eeq
The corresponding balance equation may be written as
\beq   \label{SaveNj}
\nabla_i n^i=Dn_{||}+n_{||}\theta+\nabla_i H^i=0.
\eeq

$Dw$ is substituted in the equation (\ref{convolVdT}) using
(\ref{Hibbs}). Expressing the convective derivatives $Ds_{||}$ and
$Dn_{||}$ from (\ref{SProd}) and (\ref{SaveNj}) we find
\begin{widetext}
\beq
T\nabla_i S^i - Ts_{||}\theta - T\nabla_i\sigma^i +
\sigma^i D\tau_i -
\mu_1 n_{||}\theta -\mu_1 \nabla_iH^i +w\theta +
\nabla_i Q^i -
 Q^i DV_i -\Pi^{ij}\nabla_iV_j - \frac{w}{c_p} D c_p =0.
\label{convolVdT2}
\eeq

The speed of sound $c_p$ is assumed to depend only on $n_{||}$,
and $Q^i=\Theta\sigma^i$, where the proportionality factor
$\Theta$ is to be defined. On simple rearrangement, the expression
(\ref{convolVdT2}) can be written as
\beqar                     \label{ProdEntr}
T\nabla_i S^i &=& \left\{
\Pi^{ij}-\Delta^{ij}\left(w-Ts_{||} - \mu_1 n_{||} +
\frac{wn_{||}}{c_p} \frac{\pa c_p}{\pa n_{||}}\right)+
\sigma^i \tau^j \right\} \theta_{ij}+
\nonumber\\&&
\left\{ T-\Theta \right\}\nabla_i\sigma^i +
\left\{ \mu_1-\frac{w}{c_p} \frac{\pa c_p}{\pa n_{||}}
\right\}\nabla_i H^i
+\left\{ \mathcal{L}(\Theta V_i-\tau_i)-\nabla_i\Theta
\right\}\sigma^i.
\eeqar
\end{widetext}
In the expression (\ref{ProdEntr}) we have turned from the
convective derivatives to the Lie derivatives with respect to the
vector $V^i$. This is due to the fact that the convective
derivation is concerned with changes of vector projections onto
direction $V^i$, whereas the Lie derivative does changes of the
vectors themselves. The derivatives of two types are related by
the law \beq \mathcal{L}V_i=DV_i,\qquad
\mathcal{L}\tau_i=D\tau_i+\tau^i\theta_{ij}. \eeq

Each term on the right-hand side of (\ref{ProdEntr}) includes two
factors. One of them consists of space-time derivatives, whereas
another algebraically depends on the macroscopic variables.

In equilibrium the entropy production density is equal to zero.
Since spacelike components of the flows $S^i$ and $n^i$ are
independent, each term should be put to be zero separately. This
leads to \beqar
&&\Pi^{ij}=\Delta^{ij}\left(w-Ts_{||}\right)-\sigma^i \tau^j,
\label{PIij}\\
&&T=\Theta, \\
&&\mu_1=\frac{w}{c_p} \frac{\pa c_p}{\pa n_{||}}
\label{mu1}\\
&&\nabla_i\Theta -\mathcal{L}(\Theta V_i-\tau_i)=0.
\label{convective heat}
\eeqar
The first three conditions give the connection between variables
involved in the energy-momentum tensor (\ref{decomposition}). The
last one is non-dissipative heat conduction equation.

In non-equilibrium the entropy production density is positive due
to additional contributions in the currents. It is usually
supposed that for small deviations from equilibrium these
contributions depend only linearly on the gradients of the
equilibrium quantities \cite{deGroot,Stewart}. According this
approach the right hand side of the equation (\ref{convective
heat}) becomes proportional to the non-equilibrium component in
$\sigma^i$, corresponding to an irreversible heat flow. Since this
dependence emerges in the next order of magnitude with respect to
the equilibrium values, it does not affect on the relations
(\ref{PIij})--(\ref{mu1}) (we disregard viscous effects).

In the context of the discussion regarding infinite propagation
velocity for heat signal in relativistic thermodynamics, Carter
suggested \cite{Carter6} another way to obtain the positivity of
the entropy production. This approach is based on the assumption
that heat inertia is nonzero and the energy-momentum tensor is
constructed as a purely algebraic function of independent currents
such as a particle number flow and a heat current. This statement
looks like Fok's ``physical principle'' which implies that
energy-momentum tensor components must contain the state functions
only and may not include gradients as well as coordinates in the
explicit form \cite{Fok}. Taking into account Carter's ansatz, the
relations (\ref{PIij})--(\ref{mu1}) remain valid as before and the
equation (\ref{convective heat}) transforms into
\beq   \label{ThermoEq}
\nabla_i\Theta -\mathcal{L}(\Theta V_i-\tau_i)=-Y_{ij}\sigma^j,
\eeq
where $Y_{ij}$ is a positive semidefinite resistivity matrix. The
equation (\ref{ThermoEq}) leads to a hyperbolic equation for heat
conduction automatically without using an additional
phenomenological parameter.

For our purposes the result regarding relations
(\ref{PIij})--(\ref{mu1}) is more necessary than different forms
of the heat conduction equation and it is the same at least for
the considered approaches, however the later seems to be more
consecutive in our case.

\subsection{\label{EMconserv}Phonon contribution into superfluid dynamics}

In previous section phonon gas was considered. If it is in solids,
then it is believed that phonon gas does not affect the moving of
the medium. In superfluids, phonons not only determine thermal
parameters but also change dynamic parameters of the medium. Thus
considering superfluids as closed system containing two parts,
liquid (medium) and phonons, we should assume that tensor
(\ref{decomposition}) is contained in the total energy-momentum
tensor (\ref{Tij(SF)}). Because of $\mathcal{T}^i_{j}$ is the
energy-momentum tensor of all the system it obeys the covariant
conservation law
\beq \label{Save SFE}
\nabla_i \mathcal{T}^i_{j}=0.
\eeq
At $T=0$  the energy-momentum tensor (\ref{Tij(SF)}) takes
the form
\beq \label{Tij(0)}
\mathcal{T}^i_{j}= n^i \mu_j  - P_f\delta^i_j,
\eeq
where $P_f$ is the pressure of the liquid properly, without a
phonon influence.

Using (\ref{PIij}) the energy-momentum tensor (\ref{decomposition})
can be represented in the form
\beq
T^i_j=S^i\Theta^*_j-P_{ph}\delta^i_j.
\eeq
In this case
\beqar
&&\Theta^*_j=T V_j+\tau_j,        \label{ThetaFlow}\\
&&\tau_j=\frac{T}{s_{||}}\frac{c^2}{c_p^2}\sigma_j,\\
&&P_{ph}=Ts_{||}-w.
\eeqar
Since the thermodynamic properties of superfluids
are completely determined by the phonons, and the entropy flow
vector appears in the second term of (\ref{Tij(SF)}) only,
it is reasonable to identify $\Theta^*_j$ with $\Theta_j$.

The residual part has the form (\ref{Tij(0)}) with $P_{ph}+P_f=P$,
but parameters appearing in (\ref{Tij(0)}) change their values
under the phonon influence. To obtain the phonon contribution into
the dynamic parameters one should eliminate the phonon balance
equation (\ref{SaveTij}) from the conservation law (\ref{Save
SFE}), that yields
\beqar        \label{TijFluid}
\!\!\!\!\!\!\! \nabla\!_i(n^i \mu_j-P_f\delta^i_j)\!-\!\! \left(
\frac{c^2}{c_p^2}-1 \!\right)Q^i\nabla\!_jV_i +\frac{w}{c_p}
\nabla\!_j c_p =0.
\eeqar
One transforms this expression using the condition (\ref{sf
cond}):
\beqar
n_{||}\nabla_j \left( \mu-\frac{w}{c_p} \frac{\pa c_p}{\pa n_{||}}
\right) -\nabla_j \left( P_f-\frac{w n_{||}}{c_p} \frac{\pa
c_p}{\pa n_{||}} \right)+ &&
\nonumber\\
\left(\mu H^i- \left(\frac{c^2}{c_p^2}-1\right)Q^i\right)
\theta_{ij} = 0.
&&
\label{eqnNdM}
\eeqar

The most obvious way to provide the fulfillment of this relation is
to equate the last term to zero separately.
This means that vectors $H^i$ are proportional to $\sigma^i$:
\beq      \label{HandQ}
\mu H^i= \left(\frac{c^2}{c_p^2}-1\right)T\sigma^i.
\eeq
In the residual part the expressions in parenthesis should be
identified as the chemical potential $\mu_0$ and the liquid
pressure $P_0$ at zero temperature that are connected by
$n_{||}\delta\mu_0=\delta P_0$.

Such an approach is not only possible but it allows the equation
(\ref{eqnNdM}) to be carried out for an arbitrary state equation.
Thus, at the temperature close to zero the chemical potential and
the pressure of the superfluid are determined by the equations
\beq
\mu=\mu_0+\frac{w}{c_p} \frac{\pa c_p}{\pa n_{||}}, \qquad
P_f=P_0+\frac{w n_{||}}{c_p} \frac{\pa c_p}{\pa n_{||}}.
\eeq
These expressions coincide exactly with the nonrelativistic ones
obtained for the phonon gas \cite{KhalatMono}. The additional term
in the chemical potential has the same form as $\mu_1$ introduced
in the Gibbs relation. This result could be predicted
\emph{a priori} since the physical sense of this factor
is a variation of an energy of the one liquid particle being due
to the phonon contribution  under changing a number of the
particles.

Taking into account the relations (\ref{decomposeS}),
(\ref{decomposeN}), (\ref{ThetaFlow}), and (\ref{HandQ}), two
flows, $\mu^i$ and $\Theta^i$ say, may be expressed in terms of
$n^i$ and $S^i$. Comparing them with (\ref{ABCcoef}) one can
obtain the expressions for the coefficients $A, B$, and $C$:
\beqar
A &=& - \mu T \frac{c^2/c_p^2-1}{ \mu n_{||} - s_{||} T
\left(c^2/c_p^2-1\right)},
\nonumber\\
B &=& \mu^2 \left\{ \mu n_{||} - s_{||} T \left(
c^2/c_p^2-1\right) \right\}^{-1},
\label{ABCfound}\\
C &=& \frac{T}{s_{||}} \frac{c^2}{c_p^2}\, \frac{ \left\{ \mu
n_{||} + s_{||} T \left(  1- c_p^2/c^2 \right)\right\}}{ \left\{
\mu n_{||} - s_{||} T \left(  c^2/c_p^2-1\right)\right\}}.
\nonumber
\eeqar
Since superfluid dynamics implies that $A<0$ and $B, C>0$
\cite{Carter2}, the denominator in (\ref{ABCfound}) have to be
positive. These coefficients were estimated in the work
\cite{Carter3} in which the phonon gas in superfluids is
considered from the statistical standpoint. This result can be
derived from (\ref{ABCfound}) under the condition $\mu n_{||} \gg
s_{||}T(c^2/c_p^2)$.

\section{\label{sound}Sound in superfluids at low temperature}
\subsection{\label{flat}Minkowski metrics}

Let us consider sound propagation in superfluids being in a flat
space-time at the temperature close to zero. In this temperature
range a free path of the quasi-particle increases rapidly and
equilibrium in the phonon gas has no time to be established. Under
these conditions, the hydrodynamic equations for the phonon gas
are inapplicable. Properties of this gas are described by the
kinetic equation, in which the collision integral is negligibly
small. Thus the phonon distribution function satisfies the
Liouville equation (\ref{Loivill}) with the right-hand side being
equal to zero. The Liouville equation must be supplemented by
equations of motion of the medium where the phonons are situated.
These equations are the conservation law of the particle number
flow (\ref{SaveNj}) and the irrotational condition for the
superfluid velocity (\ref{potenV}).

Solution will be sought in the form
\bsubeq    \label{SubstForSol}
\beqar
&& f=f_0+f_1e^{ik_ix^i}, \\
&& n_{||}=n_0+n_1e^{ik_ix^i}, \\
&& V^j=V_0^j+v^je^{ik_ix^i},
\eeqar
\esubeq
where $f_1,\, n_1,\, v^j$ are the small addition to the constant
equilibrium values $f_0,\, n_0,\, V_0^j$.

The equilibrium distribution function must turn the collision
integral into zero. The solution of the equation $J(x,p)=0$ for
the phonon gas (that obeys the Bose-Einstein statistics) is
well-known (see e.g. Ref.~\cite{Isihara})
\beq \label{EquilDF}
f_0=\frac{1}{e^{p_k\beta^k}-1},
\eeq
where $\beta^k=\beta V_0^k,\, \beta=c(k_B T)^{-1}$.

After linearization,
the following system of equations is derived from
(\ref{Loivill}), (\ref{SaveNj}) and (\ref{potenV})
\begin{widetext}
\bsubeq      \label{Min all}
\beqar
&& f_1 k_i \frac{\pa \mathcal{H}}{\pa p_i} +
\frac{\pa f_0}{\pa \eps} k_i \frac{\pa \mathcal{H}}{\pa p_i} p_j v^j +
\frac{\pa f_0}{\pa \eps} k_i V_0^i \left\{
\left( \frac{c^2}{c_p^2}-1 \right) \eps p_j v^j +
\frac{\eps^2}{2}\frac{\pa(c^2/c_p^2)}{\pa n_{||}} n_1
                                   \right\}=0,  \label{Min11}\\
&& k_{||} n_1 + n_0 k_j v^j + \frac{(c^2/c_p^2-1)c}{\mu} \int \eps
k_j \pi^j f_1 \d \mathcal{P} = 0,
                                                \label{Min12}\\
&& \mu k_{||} v_j = \frac{c_p^2}{c^2} \frac{\mu_0}{n_{||}}
\kappa_j n_1 + \frac{\pa}{\pa n_{||}} \left(
\frac{w}{c_p}\frac{\pa c_p}{\pa n_{||}}\right) \kappa_j n_1 +
\frac{c^3}{c_p^3}\frac{\pa c_p}{\pa n_{||}} \kappa_j \int \eps^2
f_1 \d \mathcal{P},
                                                 \label{Min13}
\eeqar
\esubeq
\end{widetext}
where $k_{||}=k_i V_0^i, \,\kappa_j=\Delta^i_j k_i$.
In the equation (\ref{Min12}) we have taken into account
the expression (\ref{HandQ}) and the fact that
$Q_i=0$ in equilibrium.

The equations (\ref{Min13}) allow to conclude that
$v_j$ is parallel to $\kappa_j$
and only one equation remains by
entering a new variable $v$ so that $v_j=v \kappa_j$.

Let us express $f_1$ from the equation (\ref{Min11}) and
substitute it in the integrals in (\ref{Min12}) and
(\ref{Min13}). It is convenient to perform the integration in
the superfluids local rest frame, marked by the condition
$V^i=\delta^i_0$.
In this frame
\begin{widetext}
\beq
f_1=\frac{\pa f_0}{\pa \eps} \frac{\eps^2}{p}
\left\{\displaystyle\left(k_{||}-\frac{c}{c_p}k\cos\vartheta\right)\frac{c}{c_p}kv\cos\vartheta -
\frac{k_{||}}{2}\frac{\pa (c^2/c_p^2)}{\pa n_{||}}\right\}
\left\{\displaystyle\frac{c}{c_p}k_{||}-k\cos\vartheta\right\}^{-1},
\eeq
where $k=(\kappa_j\kappa^j)^{1/2},\, p=(\pi_i \pi^i)^{1/2},\,$
and $\vartheta$ is the angle between the vectors $\kappa_j$
and $\pi_j$.
Having integrated over the angles we will deduce the system of equations
\bsubeq      \label{sysMink}
\beqar
&&
\displaystyle
\left\{\frac{c_p^2}{c^2}\frac{\mu_0}{n_{||}}+
       \frac{\pa}{\pa n_{||}} \left( \frac{w}{c_p}\frac{\pa c_p}{\pa n_{||}}\right)
       +\frac{c_p}{c}\frac{k_{||}}{k}\left( \frac{1}{c_p}\frac{\pa c_p}{\pa n_{||}}\right)^2 I
       \right\}n_1
-\left\{\mu k_{||} - \frac{k_{||}^2}{ck}\left(\frac{c^2}{c_p^2}-1\right)
        \frac{\pa c_p}{\pa n_{||}}I\right\}v =0,
\\&&
\displaystyle
\left\{\mu k_{||}- \frac{k_{||}^2}{ck}\left(\frac{c^2}{c_p^2}-1\right)
       \frac{\pa c_p}{\pa n_{||}}I\right\}n_1
-\left\{\mu n_0 k^2+\frac{k_{||}^3}{k}\frac{c_p}{c}
       \left(\frac{c^2}{c_p^2}-1\right)^2 I\right\}v =0,
\eeqar
\esubeq
where
\beq%
I= -2\pi c\, \ln\frac{ck_{||}+c_pk}{ck_{||}-c_pk}
  \int\frac{p^4}{(2\pi\hbar)^3}\frac{\pa f_0}{\pa \eps}\d p=
  \frac{\pi^2}{15}\frac{(k_B T)^4}{(\hbar c_p)^3}\frac{c^2}{c_p^2}
  \,\ln\frac{ck_{||}+c_pk}{ck_{||}-c_pk}.
\eeq
\end{widetext}
Only logarithmic terms are retained after
integrating over the angles. They are large in comparison with
another terms because $ck_{||} \approx c_pk$ as we will see below.

Having equated the determinant of the system (\ref{sysMink}) to
zero we will obtain the dispersion relation
\beq
\frac{k_{||}^2}{k^2}\frac{c^2}{c_p^2} = 1+
\frac{I}{\mu n_{||}}
\left\{\frac{n_{||}}{c_p}\frac{\pa c_p}{\pa n_{||}}+1-\frac{c_p^2}{c^2}\right\}^2.
\label{DispRelMink}
\eeq
The relativistic addition to the sound speed
\beq                      \label{add Cp relat}
\frac{\delta c_p}{c_p} =
\frac{I}{2\mu n_{||}}
\left\{\frac{n_{||}}{c_p}\frac{\pa c_p}{\pa n_{||}}+1-\frac{c_p^2}{c^2}\right\}^2
\eeq
differs from the nonrelativistic one \cite{KhalatMono}
\beq                      \label{add Cp class}
\delta c_p=\frac{\pi^2}{30 \hbar^3 \rho}
\left( \frac{k_B T}{c_p} \right)^4
\left\{\frac{\rho}{c_p}\frac{\pa c_p}{\pa \rho}+1\right\}^2
\ln\frac{\omega+kc_p}{\omega-kc_p},
\eeq
by the ratio of the light speed square to the sound speed one in
the braces and the production $\mu n_{||}$ instead of
nonrelativistic mass density $\rho$.

The relativistic and nonrelativistic chemical potentials are
related by $\mu=mc^2+\mu_{\scriptscriptstyle NR}$ (see e.g.
Ref.~\cite{Ehlers}). Thus the classical formula (\ref{add Cp
class}) takes into account only rest energy of the superfluids
particles whereas (\ref{add Cp relat}) includes also thermodynamic
contribution in energy.

To reduce (\ref{add Cp relat}) to (\ref{add Cp class}) it is
necessary that the conditions $c_p\ll c$ and
$\mu_{\scriptscriptstyle NR}\ll mc^2$ are performed.

To compare (\ref{add Cp relat}) and (\ref{add Cp class}) in
ultrarelativistic limit is incorrect in general, since the latter
expression is pure nonrelativistic. However it may be useful to
show how the sound speed is changed when nonrelativistic matter
transforms to ultrarelativistic one. For the latter the sound
speed $c_p=c/\sqrt{3}$ and
\beq
\frac{(\delta c_p)_{\scriptscriptstyle UR}}{(\delta
c_p)_{\scriptscriptstyle NR}}= \frac{4}{9}\frac{\rho c^2}{\mu
n_{||}}.
\eeq
This ratio demonstrates decrease of sound speed addition when
relativistic effects become essential.

\subsection{\label{RW}Robertson-Walker metrics}

In this section we consider the expanding universe with
Robertson-Walker metrics that comprises, at least in part,
superfluid matter. Similar models are regarded earlier
\cite{Silver1} for a Bose-Einstein condensate whose properties
intimately related to the superfluidity and superconductivity. In
particular it is found that the coherence length of the cosmic
Bose-Einstein condensate is equal to the Jeans wavelength. Thereby
the condensate can form gravitationally stable structures. If a
superfluid state is realized in the cosmic matter, then the
question naturally arises as to whether the excitations have an
influence on this scale. Since the Jeans scale varies directly as
the speed of sound, evaluation of $\delta c_p$ may be useful to
estimate the extent of this influence.

For the present moment we leave aside the questions of the
universe evolution under the superfluids influence and here we
study behavior of the sound speed near equilibrium.

Before to study the weakly non-equilibrium state of the phonon gas
it is necessary to examine a solution of the equations
(\ref{Loivill}), (\ref{SaveNj}) and (\ref{potenV}) at equilibrium.

The phonon gas is supposed to be at rest with respect to the
superfluids frame and the distribution function depends on only
one variable
\beq       \label{DistFo}
f_0(x,p)=f_0(p_k\beta^k),
\eeq
and has the form (\ref{EquilDF}) turning the collision integral
into zero. For the Robertson-Walker metrics with the scale factor
$a$ the dispersion relation (\ref{DispPhon}) takes the form
\[
\eps=\frac{c_p}{c}\frac{p}{a}.
\]
The kinetic equation (\ref{CartanLiouville}) leads to a law
connecting the variations of the temperature and the sound speed
in time
\beq                     \label{cp/Ta}
 \frac{\dot\beta}{\beta}-\frac{\dot
a}{a}+\frac{\dot c_p}{c_p}=0 \quad\Rightarrow\quad \frac{\beta
c_p}{a}=\mbox{const}.
\eeq

The equations of motion (\ref{SaveNj}) and (\ref{potenV})
of the liquid take the form
\bsubeq      \label{LMeq}%
\beqar%
&& \Delta_j^k\nabla\!_k \mu = \mu D V_j = 0,%
\\
&& \displaystyle          \label{na3}
 Dn_{0}+n_{0}\theta=Dn_{0}+\frac{\dot a}{a^3} n_{0} = 0,
\\&&\nonumber
\eeqar
\esubeq
where the dot denotes the time derivative.
It follows from the equations (\ref{LMeq}) that the
chemical potential is an arbitrary function of time and
the equilibrium particle number density of the liquid
varies according the law $n_0 a^3=$const.

Let us look for the weakly non-equilibrium solution in the form
(\ref{SubstForSol}). However, unlike the flat metrics case, both
the equilibrium values $f_0,\, n_0,\, V_0^j$ and the small
additions $f_1,\, n_1,\, v^j$ depend on time. These quantities
vary slower than the exponent power, i.e.
\[
\frac{\pa f_1}{\pa x^i} \ll k_i f_1,
\]
and so on.

Having repeated the procedures of the previous section, we
will obtain that
\begin{widetext}
\beq
f_1=\beta a  f_0' \frac{\eps^2}{p}
\left\{\displaystyle\left(k_{||}-\frac{c}{a c_p}k\cos\vartheta\right)
\frac{c}{a c_p}kv\cos\vartheta -
\frac{k_{||}}{2}\frac{\pa (c^2/c_p^2)}{\pa n_{||}}\right\}
\left\{\displaystyle\frac{c}{c_p}k_{||}-\frac{k}{a}\cos\vartheta\right\}^{-1},
\eeq
\end{widetext}
and integration over the angles gives the following result
\beqaw
I&=&-\frac{2\pi\beta}{a^5}\,
\ln\frac{ack_{||}+c_pk}{ack_{||}-c_pk}
  \int\frac{p^4}{(2\pi\hbar)^3}f_0'\d p
\\  &=&
  \frac{\pi^2}{15}\frac{(k_B T)^4}{(\hbar c_p)^3}\frac{c^2}{c_p^2}
  \,\ln\frac{ack_{||}+c_pk}{ack_{||}-c_pk},
\eeqaw
where the prime denotes the derivative with respect to the argument.
To write down the dispersion relation for the Robertson-Walker metrics,
one should add $a^2$ to the numerator of the right hand side of
the expression (\ref{DispRelMink}).

In view of (\ref{cp/Ta}) and (\ref{na3}) $\delta c_p \propto 1/\mu
a$ (logarithmic dependence is neglected, since it related  with
dispersion of phonons). Further consideration is caused  by
behavior of the chemical potential, that can not be determined
without additional assumptions. As possible models one can
consider the simple equations of state that allows to examine the
evolution of $\delta c_p$.

(i) Linear barotropic equation of state takes the form
$P_0=(c_p^2/c^2)\mu_0 n_{||}$ for superfluid matter at zero
temperature where $c_p=\mbox{const}$. Integrating the relation
(\ref{def Cp}) one obtains $\mu_0\propto n_{||}^{c_p^2/c^2}$ and
therefore $\delta c_p \propto a^{3c_p^2/c^2-1}$.

In the ultrarelativistic limiting case the addition to the sound
speed is independent from the scale factor $a$. Thus the sound
speed is constant but does not coincide with one at zero
temperature. For the nonrelativistic limiting case $\delta c_p
\propto a^{-1}$ and universe inflationary implies decay of the
phonon contribution  into the sound speed.

(ii) Polytropic equation of state $P_0=Kn_{||}^\gamma$ ($\gamma>1$
is supposed in judgements) in combination with the relations
(\ref{def Cp}) permits to obtain the expressions:
\beqaw
& \displaystyle
\frac{c_p^2}{c^2}=\frac{K\gamma
n_{||}^{\gamma-1}}{mc^2+K\gamma
n_{||}^{\gamma-1}/(\gamma-1)}, & \\
& \displaystyle
\mu_0=mc^2+\frac{\gamma}{\gamma-1}Kn_{||}^{\gamma-1},
&
\eeqaw
which is similar to the expressions for nonrelativistic polytropes
\cite{Weinberg} because of their temperature independence.

In the nonrelativistic area ($mc^2\gg Kn_{||}^{\gamma-1}$) the
sound speed $c_p\propto a^{-3(\gamma-1)/2}$ and $\delta c_p
\propto a^{-1}$. One can see that for $\gamma>5/3$ the sound speed
decreases slower then the phonon addition. In general, the same
behavior also takes place in the relativistic area ($mc^2\sim
Kn_{||}^{\gamma-1}$) especially for the large values of $a$. When
$mc^2\ll Kn_{||}^{\gamma-1}$ and the scale factor is not so large,
then $c_p=\mbox{const}$ and $\delta c_p \propto a^{3\gamma-4}$.
The value $\gamma=4/3$ leads us again to the ultrarelativistic
limits described in (i).

\section{Conclusion}
In this paper we introduced the principle of constructing
macroscopic flows of the phonon gas in superfluids using the
phonon distribution function. These definitions can be applied to
arbitrary quasi-particles without any changes. Nevertheless we
restrict ourselves by consideration of the phonon gas. In this
case there is no dependence on the quasi-particle energy in
$\overline{g}^{ij}$ so this tensor can be used as an effective
second metrics.

This allowed us to apply the well developed technique to derive
the macroscopic balance equations for the phonon gas, which is the
part of the hydrodynamic conservation laws. Based on this
statement the phonon contribution in superfluid dynamics was
obtained.

It is worth noting that using the expression (\ref{def Tij})
instead of (\ref{old Tij}), one can draw an analogy with the works
\cite{Gordon,Nelson,Stone1}(see also references therein) where a
distinction between momentum and pseudomomentum is discussed. This
problem is considered for dielectric media and electromagnetic
waves in \cite{Gordon,Nelson} and for acoustic waves in
\cite{Stone1}. In these papers the mentioned problem is studied
for an individual optical or acoustic vibrations. In contrast to
such approach, the tensors (\ref{old Tij}) and (\ref{def Tij}) do
not contain parameters of individual phonons but describe the
quasi-particle ensemble. Therefore, it is reasonable to consider
concepts of momentum and pseudomomentum in order to compare their
applicability to description  of macroscopic objects.

Obtained in this paper kinetic equation was solved directly to
determine the addition to the speed of sound at temperature close
to zero.

The result for the flat space-time is the relativistic
generalization of the classical Khalatnikov's one \cite{Khalat},
whereas results from Sec.~\ref{RW} should be reviewed from the
other standpoint. They are certainly preliminary and estimating
ones, however they demonstrate that including phonons into
cosmological models can be used to fit necessary parameters
properly.

The present research was carried out for the low temperature
quasi-particle gas, when purely phonon processes dominate, whereas
contributions of quasi-particles corresponding to another parts of
the spectrum are neglected. To develop the similar theory for
another sorts of quasi-particles, such as rotons, special
techniques are needed because of the nonlinear dependence between
the energy and momentum of a quasi-particle.

\begin{acknowledgments}
This paper was supported by the Russian Program of Support of the
Leading Scientific Schools (grant
HL\hspace{-.2em}L\hspace{-.25em}I-1789.2003.02). The author would
like to thank A.~B.~Balakin for stimulating conversations and
M.~Kh.~Brenerman for help in preparing the paper.
\end{acknowledgments}


\newpage
\begin{thebibliography}{}
\bibitem{Israel}
W.~Israel, Phys.~Lett. \textbf{86A}, 79 (1981); \textbf{92A}, 77 (1982).%
\bibitem{KhalatLeb}
I.~M.~Khalatnikov and V.~V.~Lebedev, Phys.~Lett. \textbf{91A}, 70 (1982).
\bibitem{LebKhal}
V.~V.~Lebedev and I.~M.~Khalatnikov, Zh. Eksp. Teor. Fiz. \textbf{83}, 1601 (1982)
[Sov.~Phys.~JETP \textbf{56}, 923 (1982)].
\bibitem{KhalatCarter1}
B.~Carter and I.~M.~Khalatnikov, Phys.~Rev.~D \textbf{45}, 4536 (1992).
\bibitem{Carter2}
B.~Carter and I.~M.~Khalatnikov, Ann. Phys. (N.Y.) \textbf{219}, 243 (1992).
\bibitem{Carter6}
B.~Carter, in \emph{A Random Walk in Relativity and
Cos\-mo\-lo\-gy}, (Proc. Vadya Raychaudhuri Festschrift,
I.A.G.R.G. 1983), edited by N.~Dadhich, J.~Krishna Rao,
J.~V.~Narlikar, C.~V.~Vishveshwara, (Wiley Easter, Bombey, 1985),
pp.~48--62.
\bibitem{Carter3}
B.~Carter and D.~Langlois, Phis.~Rev.~D \textbf{51}, 5855 (2000).
\bibitem{Comer}
G.~L.~Comer, D.~Langlois, and L.~M.~Lin, Phys.~Rev.~D \textbf{60}, 104025 (1999).
\bibitem{Lindblom}
L.~Lindblom and G.~Mendell, Phys.~Rev.~D \textbf{61}, 104003 (2000).
\bibitem{Yakovlev1}
P.~Haensel, K.~P.~Levenfish, and D.~G.~Yakovlev, Astron. Astrophys. \textbf{357}, 1157 (2000);
\textbf{372}, 130 (2001).
\bibitem{AnderComer1}
N.~Andersson and G.~L.~Comer, Class.~Quant.~Grav. \textbf{18}, 969 (2001);
Phys.~Rev.~Lett. \textbf{87}, 241101 (2001).
\bibitem{Volovik2}
G.~E.~Volovik, Physics Reports \textbf{351}, 195 (2001).
\bibitem{Volovik3}
U.~R.~Fischer and G.~E.~Volovik, Int.~J.~Mod.~Phys.~D \textbf{10}, 57 (2001)
\bibitem{Visser}
C.~Barcelo, S.~Liberati, and M.~Visser, Class. Quant. Grav. \textbf{18}, 1137 (2001)
\bibitem{Khalat}
I.~M.~Khalatnikov, Zh. Eksp. Teor. Fiz. \textbf{23}, 8 (1952).
A comprehensive description of the results of this work can be found in
Ref.~\cite{KhalatMono}
\bibitem{deGroot}
S.~R.~de~Groot, W.~A.~van~Leeuwen, and Ch.~G.~van Weert, \emph{Relativistic Kinetic
Theory} (North~Holland, Amsterdam, 1980).
\bibitem{KhalatMono}
I.~M.~Khalatnikov, \emph{An Introduction to the Theory of Superfluidity}
(Benjamin, New York, 1965).
\bibitem{LL}
L.~D.~Landau and E.~M.~Lifshitz, \emph{Course of Theoretical Physics,
Vol.~10: Physical Kinetics} (Pergamon, Oxford, 1981).
\bibitem{Sing}
J.~L.~Synge, \emph{Relativity: The General Theory} (North~Holland, Amsterdam, 1960).
\bibitem{unruh}
W.~G.~Unruh, Phys.~Rev.~D \textbf{51}, 2827 (1995).
\bibitem{FirstSound}
This corresponds to usual sound waves, named also the first sound.
The second sound corresponding to excitations in the
quasi-particle gas, is not concerned here.
\bibitem{Stewart}
J.~M.~Stewart, \emph{Non-Equilibrium Relativistic Kinetic Theory}
(Springer, New York, 1971).
\bibitem{LL2}
L.~D.~Landau and E.~M.~Lifshitz, \emph{Course of Theoretical Physics,
Vol.~2: The Classical Theory of Fields} (Pergamon, Oxford, 1975).
\bibitem{Klim}
Y.~A.~Klimontovich, Zh. Eksp. Teor. Fiz. \textbf{37}, 735 (1959)
[Sov.~Phys.~JETP \textbf{37}, 524 (1960)].
\bibitem{Vlasov}
A.~A.~Vlasov, \emph{Statistical Functions of Distribution} (Na\-u\-ka, Mos\-cow, 1966)
[in Russian].
\bibitem{Fok}
V.~A.~Fok, \emph{The Theory of Space, Time and Gravitation}
(Pergamon, London, 1959).
\bibitem{Isihara}
A.~Isihara, \emph{Statistical Physics} (Acad.~Press, New York, 1971).
\bibitem{Ehlers}
J.~Ehlers, in \emph{General Relativity and Cosmology}, edited by
B.~K.~Sachs (Academic, New York, 1971).
\bibitem{Silver1}
M.~P.~Silverman and R.~L.~Mallett, Class.~Quant.~Grav.
\textbf{18}, L103 (2001).
\bibitem{Weinberg}
S.~Weinberg, \emph{Gravitation and Cosmology. Principles and
applications of the theory of relativity} (J.~Wiley and Sons, New
York, 1972).
\bibitem{Gordon}
J.~P.~Gordon, Phys.~Rev.~A \textbf{8}, 14 (1973).
\bibitem{Nelson}
D.~F.~Nelson Phys.~Rev.~A \textbf{44}, 3985 (1991).
\bibitem{Stone1}
M.~Stone Phys.~Rev.~E \textbf{62}, 1341 (2000).
\end{thebibliography}
\end{document}